\documentclass{JHEP}
 \usepackage{epsfig}

 \title{
  Single charged Higgs boson production at next generation linear
  colliders}

 \author{Shinya Kanemura\\
  Physics and Astronomy Department, Michigan State University\\
  East Lansing, MI 48824--1116, USA}

 \author{Stefano Moretti\\
  Theory Division, CERN, CH--1211 Gen\`eve 23, Switzerland}

 \author{Kosuke Odagiri\\
  Theory Group, KEK, 1--1 Oho, Tsukuba, Ibaraki 305--0801, Japan}

 \abstract{
  We study the single production of charged Higgs bosons in $e^+e^-$
collisions, chiefly in the Minimal Supersymmetric Standard Model. Our
analysis complements foregoing studies of the pair production channel
especially in regions where the kinematic constraint suppresses pair
production. We present cross sections relevant to experiments at the next
generation linear colliders and some brief discussions of their
phenomenology.
  Our analysis shows that the single production is a useful alternative
channel for studying the phenomenology of charged Higgs bosons, and there
are regions of parameter space where it is accessible beyond the kinematic
limit for pair production.
 }

 \keywords{Beyond Standard Model, Supersymmetric Models, Higgs Physics}

 \preprint{CERN--TH/2000--347\\
           KEK--TH--729\\
           MSUHEP--00911\\
           RAL--TR--2000--037}

\begin{document}

 \section{Introduction}
  Charged Higgs bosons $H^\pm$ are a cornerstone of
beyond-the-Standard-Model phenomenology. They arise as a prediction of the
supersymmetric extensions of the Standard Model (SM) from purely
theoretical requirements of consistency. Their discovery and the
confirmation of their properties will be a significant step towards a full
understanding of electroweak symmetry breaking.

  Some substantial effort has thus been channelled into the evaluation of
their phenomenology at future colliders, and over the past few years the
situation with respect to their discovery potential has become more clear.

  At hadron colliders \cite{LHC_search}, especially with the luminosity of
the LHC, charged Higgs bosons below the top quark mass are expected to be
produced abundantly in the decay of top quarks. When their mass is near or
greater than the top quark decay threshold, their discovery prospects are
further hindered by the falling structure function but they will be
produced mainly in the `$tb$-fusion' process, namely the parton level
process $gb\to tH^-$ and the charge conjugate. There have recently been
attempts \cite{gg_to_tbH} to connect the two channels together by looking
at a generic subprocess $gg\to t\bar bH^-$ which includes a component of
the latter process that is leading order at large transverse momentum of
the `spectator' bottom quark. In both cases the most promising decay mode
is presumably the $H^-\to\tau^-\bar\nu$ mode \cite{ko_tau_nu}. It is
claimed that relatively heavy charged Higgs bosons, of mass up to 1 TeV,
can be probed using this mode, especially at large $\tan\beta$, although a
full experimental simulation similar to that presented in
\cite{Les_Houches} will be essential to test this claim and to draw the
discovery contour. Here $\tan\beta$ is as usual defined as the ratio of
the vacuum expectation values of the two Higgs doublets. Several other
production channels have also been considered \cite{other_search}.

  At future electron-positron linear colliders (LCs), the dominant 
production process,
if kinematically allowed, is the pair production process \cite{ee_pair}:
  \begin{equation}\label{proc_pair}
    e^-e^+ \to H^+H^-.
  \end{equation}
  The production rate depends only on the charged Higgs mass and the
process provides a hallmark channel through which we can study $H^\pm$
phenomenology \cite{pair_phenom}.

  When the charged Higgs mass is near or above half the centre-of-mass
energy the phenomenology is much less well understood. There has lately
been some interest in the associated production mode with $W^\pm$
\cite{zhu_HW, sk_HW, hollik_HW}:
  \begin{equation}\label{proc_whi}
     e^-e^+ \to H^\pm W^\mp,
  \end{equation}
  although the extent to which this channel could contribute towards the
study of charged Higgs phenomenology is not clear. In models with a 
multi-Higgs-doublet structure, the process is loop-induced in the massless 
electron limit, and therefore the cross section is suppressed. We also note
that no analysis of the SM background is available so far.

  Other channels that have been studied in the literature are the heavy
quark associated production mode \cite{ee_bbWH}:
  \begin{equation}\label{proc_bbwhi}
     e^-e^+ \to b\bar b W^\pm H^\mp,
  \end{equation}
  and the $\tau\nu_\tau$ associated production mode \cite{ee_taunuH}:
  \begin{equation}\label{proc_taui}
     e^-e^+ \to H^+\tau^-\bar\nu_\tau, H^-\tau^+\nu_\tau.
  \end{equation}

  The purpose of this paper is to study the above processes in greater
detail and to collect them together with many other single production channels
of charged Higgs bosons in order to complement the pair production channel
both above and below its kinematic threshold.
  Although our method is applicable generally in principle, we have adopted
the Minimal Supersymmetric Standard Model (MSSM) for calculating Higgs
masses and mixings. The MSSM being a model with a decoupling structure in
the Higgs sector, the resulting cross sections are small compared to cases
where, for example, extra resonances are available. 
However, it is beyond our intention to study models other than the 
MSSM in the present paper.

 \section{Production processes}
  We consider the following fourteen processes:
  \begin{eqnarray}
   e^-e^+ &\to& \tau^-\bar\nu_\tau H^+, \tau^+\nu_\tau H^-\label{proc_tau} \\
   e^-e^+ &\to& \bar tbH^+, t\bar bH^-              \label{proc_tb}  \\
   e^-e^+ &\to& W^\mp H^\pm \mathrm{(one\ loop)}    \label{proc_wh}  \\
   e^-e^+ &\to& e^-\bar\nu H^+, e^+\nu H^-
                \mathrm{ (one\ loop)}               \label{proc_enu} \\
   e^-e^+ &\to& Z^0W^\mp H^\pm                      \label{proc_zwh} \\
   e^-e^+ &\to& h^0W^\mp H^\pm                      \label{proc_lwh} \\
   e^-e^+ &\to& H^0W^\mp H^\pm                      \label{proc_bwh} \\
   e^-e^+ &\to& A^0W^\mp H^\pm                      \label{proc_awh} \\
   e^-e^+ &\to& e^-e^+W^\mp H^\pm                   \label{proc_eewh}\\
   e^-e^+ &\to& \nu_e\bar\nu_eW^\mp H^\pm           \label{proc_nnwh}\\
   e^-e^+ &\to& e^-\bar\nu_eZ^0 H^+, e^+\nu_eZ^0H^- \label{proc_enzh}\\
   e^-e^+ &\to& e^-\bar\nu_eh^0 H^+, e^+\nu_eh^0H^- \label{proc_enlh}\\
   e^-e^+ &\to& e^-\bar\nu_eH^0 H^+, e^+\nu_eH^0H^- \label{proc_enbh}\\
   e^-e^+ &\to& e^-\bar\nu_eA^0 H^+, e^+\nu_eA^0H^-.\label{proc_enah}
  \end{eqnarray}
  The Feynman graphs corresponding to the above are shown in figure
\ref{feynman_graphs}.

  \FIGURE[p]{
  \epsfig{file=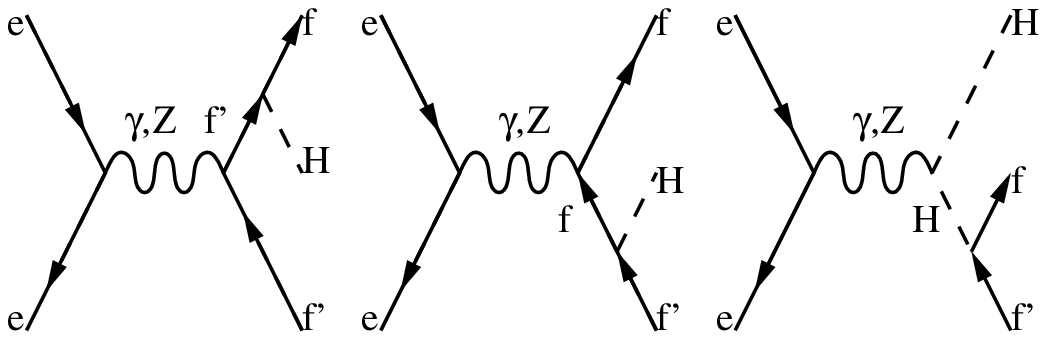,height=4cm}\\[1cm]
  (a) processes (\ref{proc_tau}) and (\ref{proc_tb})\\[1cm]
  \epsfig{file=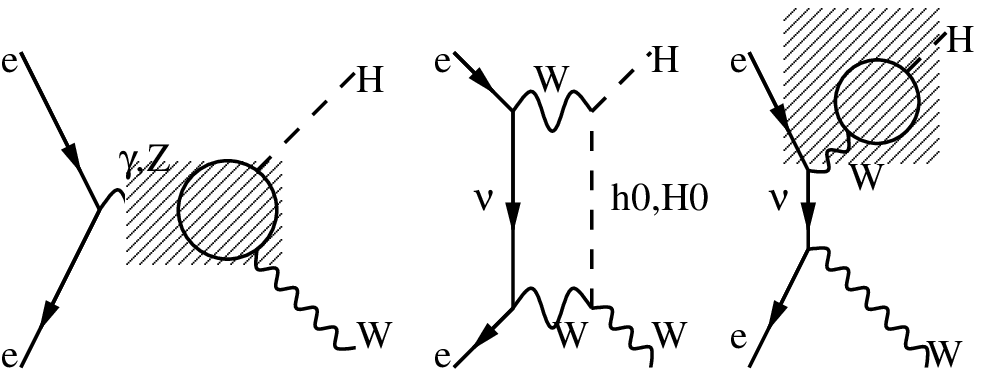,height=3.8cm}\\[1cm]
  (b) process (\ref{proc_wh}). Shaded circles represent one-loop
  contributions.\\[1cm]
  \epsfig{file=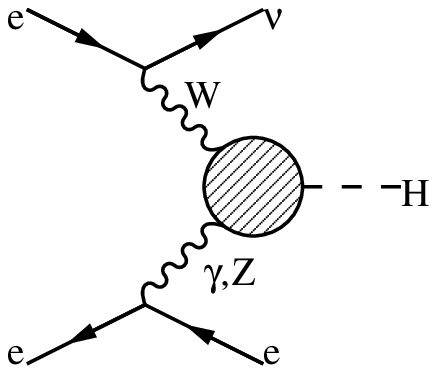,height=4cm}\\[1cm]
  (c) process (\ref{proc_enu}). $W^\pm-H^\pm$ mixing is
  taken into account by using the renormalised couplings. We neglect the
  box diagram.
  }
  \FIGURE[p]{
  \epsfig{file=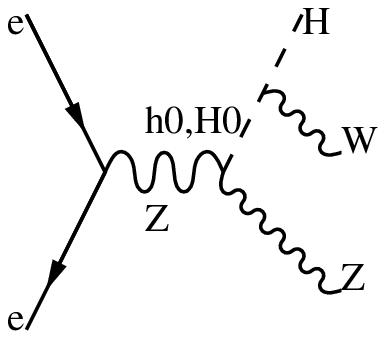,height=4cm}\\[1cm]
  (d) process (\ref{proc_zwh})\\[1cm]
  \epsfig{file=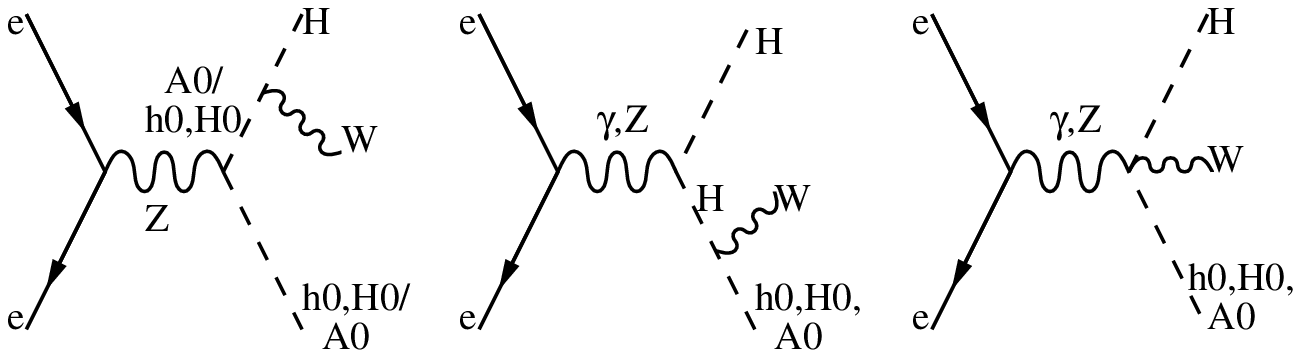,height=4cm}\\[1cm]
  (e) processes (\ref{proc_lwh})--(\ref{proc_awh})\\[1cm]
  \epsfig{file=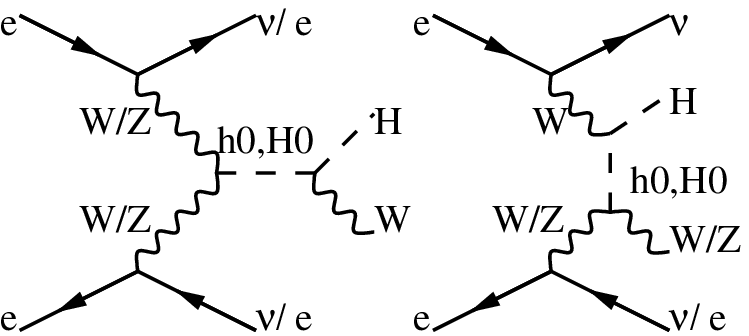,height=3.6cm}\\
  (f) processes (\ref{proc_eewh})--(\ref{proc_enzh})
  }
  \FIGURE[p]{
  \epsfig{file=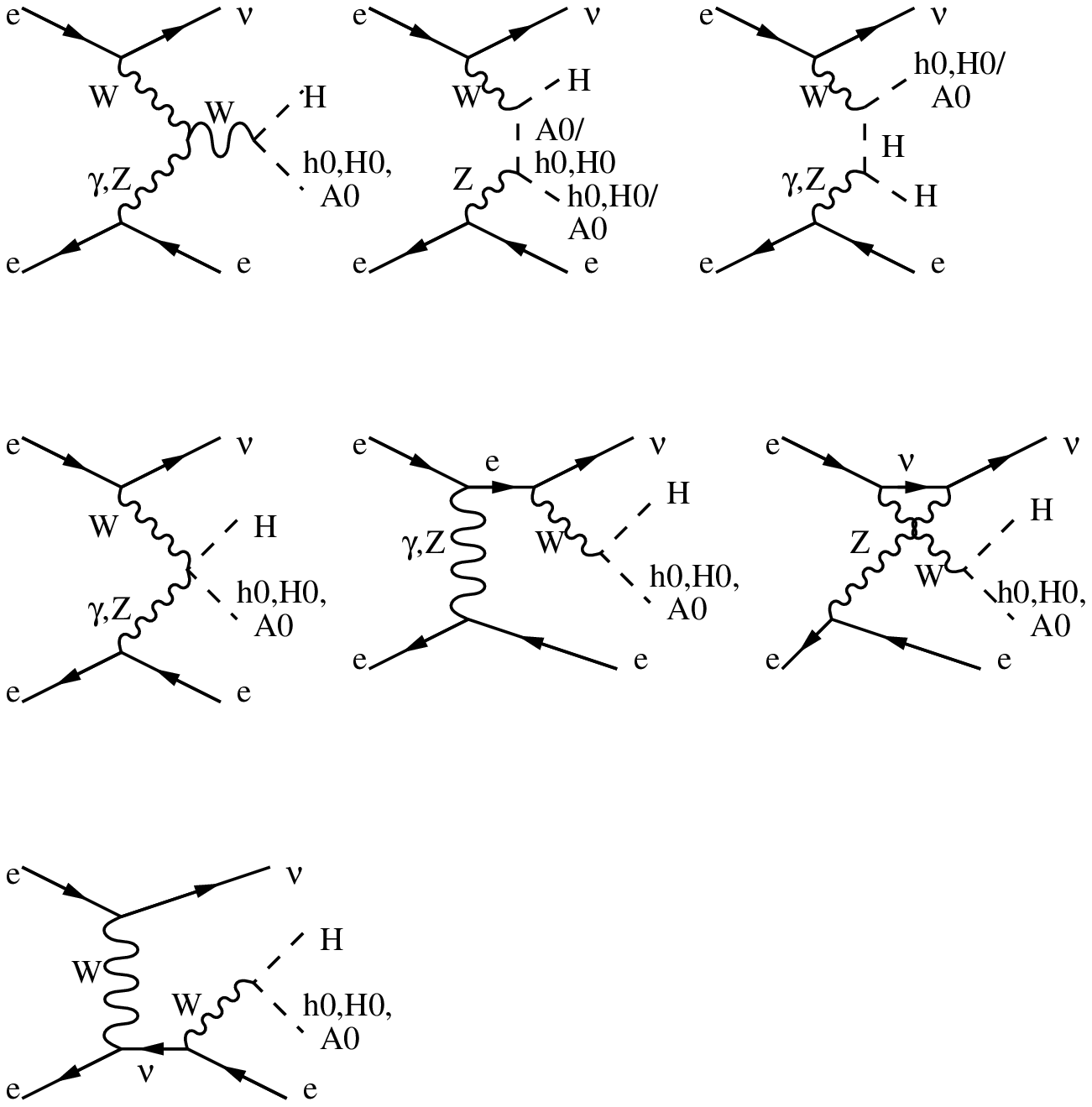,height=15cm}\\[1cm]
  (g) processes (\ref{proc_enlh})--(\ref{proc_enah})\\[1cm]
  \caption{Feynman diagrams for the single $H^\pm$ production
  processes. In all diagrams $H$ stands for $H^\pm$ and $W$ stands for
  $W^\mp$, the charge being dictated by charge conservation in the 
  diagram concerned. Charge conjugated diagrams have been omitted from
  (b), (c) and (f) for simplicity.}
  \label{feynman_graphs}
  }

  For all tree-level processes, the matrix elements were calculated both by
hand, for select set of diagrams in order to simplify the calculation,
using the usual trace method, and by using the helicity amplitude formalism
of \cite{helicity_HZ}. The two sets of results agree up to the numerical
precision employed. We present the matrix element squared for the 
dominant tree-level signal 
processes (\ref{proc_tau}) and (\ref{proc_tb}) in the
appendix.

  All processes were calculated at leading order only.
  For the 2HDM parameters, we adopted the MSSM throughout.
  For the SM parameters we adopted the following:
  $m_b=4.25$ GeV, $m_t=175$ GeV, $m_e=0.511$ MeV, $m_\tau=1.78$ GeV,
$m_\nu=0$, $M_W=80.23$ GeV, $\Gamma_W=2.08$ GeV, $M_Z=91.19$ GeV,
$\Gamma_Z=2.50$ GeV, $\sin^2\theta_W=0.232$. The top quark width $\Gamma_t$
was evaluated at leading order for each value of $M_{H^\pm}$ and
$\tan\beta$. Neutral and charged Higgs masses were calculated for given
values of $M_{A^0}$ and $\tan\beta$ using the HDECAY package \cite{hdecay},
with the SUSY masses, the trilinear couplings and the Higgsino mass
parameter $\mu$ being set to 1 TeV. The charged Higgs boson width
$\Gamma_{H^\pm}$ was evaluated at leading order.

  The loop-induced processes (\ref{proc_wh}) and (\ref{proc_enu}) were
adapted from \cite{sk_HW, sk_WZdec}.
In the one-loop analyses, we
assumed that the superpartners are sufficiently heavy to decouple, so that
only the heavy quark loops and Higgs--gauge loops are included. We
introduced counter terms from $WH$ and $wH$ mixings ($w$ represents the
Nambu-Goldstone boson) \cite{capdequi} and used the renormalised coupling
to account for the mixing. Details of the calculation are shown in
\cite{sk_HW}. One of the renormalisation conditions is that the $WH$ mixing
is zero for onshell $H^\pm$, such that the third diagram of figure
\ref{feynman_graphs}(b) is effectively zero. The cross sections are
qualitatively consistent with other calculations \cite{zhu_HW, hollik_HW},
although we note that this renormalisation is not performed in
\cite{zhu_HW}.

  The $HWZ$ and $HW\gamma$ vertices, which enter into the process
(\ref{proc_enu}), have been calculated in \cite{sk_WZdec} and
\cite{rizzomendez} for the $HWZ$ vertex and in \cite{haber} for both
vertices. In our analysis, we improved on their calculations by
renormalising the $WH$ and $wH$ mixings according to the calculation in
\cite{sk_HW} and \cite{hollik_HW}.
 Again, the superpartners were assumed to be heavy. In addition, for
process (\ref{proc_enu}), we dropped the bosonic loop contributions in
order to save time during numerical computation. This does not affect the
results, as the contributions from Higgs--gauge loop diagrams are small in
parameter regions where these vertices are substantial. We have explicitly
verified this statement numerically. The total cross sections were
evaluated using these vertices with the helicity amplitude method.

  In the results which we present, the charge conjugate subprocesses are
not included. Our results correspond to the production of either an $H^+$
or an $H^-$ (except for the pair production process). This is in order to
avoid double counting final states. For instance, if we consider
process (\ref{proc_tau}), the final state can be $\tau^-\bar\nu_\tau\
 \tau^+\nu_\tau$.
 Far from the kinematic threshold for pair
production the `total' single charged Higgs production for a final state
$XH^\pm$ may be given by:
  \begin{equation}
   \sigma(XH^\pm)=\sigma(XH^+)+\sigma(XH^-)-BR(H^\pm\to X)\sigma(H^+H^-).
  \end{equation}
  Near the kinematic threshold we can not treat the processes consistently
using this formula without specifying the particular final state, and this
would limit the generality of our approach.
 The total cross section is twice that presented here in regions where pair
production is forbidden, and nearly equal to that presented here in the
limit where the branching ratio $BR(H^\pm\to X)$ tends to one.

 \section{Production cross sections}
  We present the cross sections as functions of the charged Higgs boson
mass $M_{H^\pm}$ at collider energies of 500 GeV and 1000 GeV and 4
different values of $\tan\beta$, 1.5, 7, 30 and 40. These are shown in
figures \ref{rate_tau} to \ref{rate_enlh}. If we assume, for instance, an
integrated luminosity of 500 fb$^{-1}$  \cite{nlc_params},
$10^{-5}$ pb corresponds to 5 events per year before acceptance cuts and
background reduction. We do not discuss the background reduction procedure
in detail in this study, and $10^{-5}$ pb is taken naively as the threshold
of the `relevance' of the process to the study of charged Higgs production
at LC. We emphasise that this is not intended in any way as a threshold of 
detectability, or even visibility, as the evaluation of such thresholds 
would require jet simulations and machine-dependent considerations which 
are clearly beyond the scope of this current study.

  \FIGURE[p]{
  \epsfig{file= plots/eetnhpm_500.ps,width=5cm,angle=90}
  \epsfig{file=plots/eetnhpm_1000.ps,width=5cm,angle=90}\\
  (a) Process (\ref{proc_tau})\\[1cm]
  \epsfig{file= plots/eebthpm_500.ps,width=5cm,angle=90}
  \epsfig{file=plots/eebthpm_1000.ps,width=5cm,angle=90}\\
  (b) Process (\ref{proc_tb})\\
  \caption{Total cross sections for the tau and the heavy quark
  associated production channels.}
  \label{rate_tau}
  }

  \FIGURE[p]{
  \epsfig{file= plots/eewhpm_500.ps,width=5cm,angle=90}
  \epsfig{file=plots/eewhpm_1000.ps,width=5cm,angle=90}\\
  \caption{Total cross sections for the $W^- H^+$ associated
  production process (\ref{proc_wh}).}
  \label{rate_wh}
  }

  \FIGURE[p]{
  \epsfig{file= plots/eeenhpm_500.ps,width=5cm,angle=90}
  \epsfig{file=plots/eeenhpm_1000.ps,width=5cm,angle=90}\\
  \caption{Total cross sections for the loop induced vector fusion process
  (\ref{proc_enu}).}
  \label{rate_enu}
  }

  \FIGURE[p]{
  \epsfig{file= plots/eezhpmw_500.ps,width=5cm,angle=90}
  \epsfig{file=plots/eezhpmw_1000.ps,width=5cm,angle=90}\\
  \caption{Total cross sections for the $Z^0W^- H^+$ associated
  production process (\ref{proc_zwh}).}
  \label{rate_zwh}
  }

  \FIGURE[p]{
  \epsfig{file= plots/eeh02hpmw_500.ps,width=5cm,angle=90}
  \epsfig{file=plots/eeh02hpmw_1000.ps,width=5cm,angle=90}\\
  (a) Process (\ref{proc_lwh})\\[1cm]
  \epsfig{file= plots/eeh01hpmw_500.ps,width=5cm,angle=90}
  \epsfig{file=plots/eeh01hpmw_1000.ps,width=5cm,angle=90}\\
  (b) Process (\ref{proc_bwh})\\[1cm]
  \epsfig{file= plots/eeh03hpmw_500.ps,width=5cm,angle=90}
  \epsfig{file=plots/eeh03hpmw_1000.ps,width=5cm,angle=90}\\
  (c) Process (\ref{proc_awh})\\[1cm]
  \caption{Total cross sections for the neutral Higgs associated
  production channels.}
  \label{rate_lwh}
  }

  \FIGURE[p]{
  \epsfig{file= plots/eeeehpmw_500.ps,width=5cm,angle=90}
  \epsfig{file=plots/eeeehpmw_1000.ps,width=5cm,angle=90}\\
  (a) Process (\ref{proc_eewh})\\[1cm]
  \epsfig{file= plots/eennhpmw_500.ps,width=5cm,angle=90}
  \epsfig{file=plots/eennhpmw_1000.ps,width=5cm,angle=90}\\
  (b) Process (\ref{proc_nnwh})\\[1cm]
  \epsfig{file= plots/eeenzhpm_500.ps,width=5cm,angle=90}
  \epsfig{file=plots/eeenzhpm_1000.ps,width=5cm,angle=90}\\
  (c) Process (\ref{proc_enzh})\\[1cm]
  \caption{Total cross sections for the vector fusion mediated gauge boson
  associated production channels.}
  \label{rate_eewh}
  }

  \FIGURE[p]{
  \epsfig{file= plots/eeenh02hpm_500.ps,width=5cm,angle=90}
  \epsfig{file=plots/eeenh02hpm_1000.ps,width=5cm,angle=90}\\
  (a) Process (\ref{proc_enlh})\\[1cm]
  \epsfig{file= plots/eeenh01hpm_500.ps,width=5cm,angle=90}
  \epsfig{file=plots/eeenh01hpm_1000.ps,width=5cm,angle=90}\\
  (b) Process (\ref{proc_enbh})\\[1cm]
  \epsfig{file= plots/eeenh03hpm_500.ps,width=5cm,angle=90}
  \epsfig{file=plots/eeenh03hpm_1000.ps,width=5cm,angle=90}\\
  (c) Process (\ref{proc_enah})\\[1cm]
  \caption{Total cross sections for the vector fusion mediated neutral
  Higgs associated production channels.}
  \label{rate_enlh}
  }

  For the sake of comparison we also present the cross sections of the
on-shell pair production mode (\ref{proc_pair}) in figure \ref{rate_pair}.

  \FIGURE[p]{
  \epsfig{file=plots/eehphm.ps,width=8cm,angle=90}\\
  \caption{Total cross sections for the pair production channel
  (\ref{proc_pair}).}
  \label{rate_pair}
  }

  In the allowed kinematic ranges, the $\tau\nu_\tau$ and $tb$ associated
production processes (\ref{proc_tau}) and (\ref{proc_tb}), shown in figure
\ref{rate_tau}, are dominated by pair production. When $M_{H^\pm}$ is
small, process (\ref{proc_tb}) also has a large contribution from top pair
production followed by the decay of one of the top quarks into $bH^\pm$,
which explains the rise of the cross section below 175 GeV. Beyond the
kinematic limit for pair production, which occurs at
$M_{H^\pm}\sim\sqrt{s}/2$, the cross sections still exceed $10^{-5}$ pb for
some values of $\tan\beta$. Process (\ref{proc_tau}) is enhanced for large
$\tan\beta$ whereas process (\ref{proc_tb}) is enhanced for both large and
small values of $\tan\beta$ and the minimum is at
$\tan\beta=\sqrt{m_t/m_b}\sim7$.

  Figure \ref{rate_wh} shows the rate for the loop-induced $W^\pm H^\mp$
associated production process (\ref{proc_wh}). When $\tan\beta$ is small,
the top Yukawa coupling is enhanced and the cross section is large. Hence
this channel complements process (\ref{proc_tau}). We note that the
$\tan\beta$ dependence of the signal rate for this process is $\sim m_t^4
\cot^2\beta$ at small $\tan\beta$ and $\sim m_b^4\tan^2\beta$ at very large
$\tan\beta$. Hence the bottom Yukawa contribution is suppressed.
  The cross section remains large beyond the pair production kinematic
limit. The peak in the cross section near 200 GeV corresponds to the
threshold $m_t+m_b\sim M_{H^\pm}$, after which the cross section falls
slowly up to the kinematic limit at $M_{H^\pm} \sim\sqrt{s}-M_{W^\pm}$.

  We note that at $\tan\beta\sim7$, both the $\tau^-\bar\nu_\tau H^+$ mode and
the $W^\pm H^\mp$ mode have cross sections near $10^{-5}$ pb at $\sqrt{s}=
500$ GeV and $M_{H^\pm}>250$ GeV. Thus there are always regions in
$M_{H^\pm}$ where charged Higgs bosons can be produced at a 500 GeV machine
even when the pair production channel is unavailable. However, we should
bear in mind that the detectability of such processes needs more realistic
simulations in order to quantify the effect of the decay, the fragmentation
and detector resolutions.

  In figure \ref{rate_enu} we show the rate for the loop induced vector
boson fusion process (\ref{proc_enu}). The overall rate is small at both
500 GeV and 1 TeV, but there is an interesting $\tan\beta$ dependence of
the cross section where, near the $t\bar b\to H^+$ kinematic threshold, the
rates are enhanced for large $\tan\beta$. As with the other vector fusion
induced processes which we present later, the cross section is larger at 1
TeV because of the `$t$-channel' vector boson propagators.

  Figure \ref{rate_zwh} shows the rate for the $W^\pm Z^0H^\mp$ associated
production process (\ref{proc_zwh}). The cross section is small for all
parameter values that we are considering. This is because the amplitude is
proportional to $\sin(\beta-\alpha)\cos(\beta-\alpha)$ and this is
suppressed in the MSSM as, in the decoupling limit when $M_{H^\pm}$
becomes large, $\cos(\beta-\alpha)$ becomes small. There is also a
cancellation between the $h^0$ and $H^0$ mediated diagrams.

  The situation with process (\ref{proc_zwh}) is in good contrast with 
$(h^0/H^0/A^0) W^\pm H^\mp$ associated production,
(\ref{proc_lwh}) -- (\ref{proc_awh}), shown in figures \ref{rate_lwh}.  
Process (\ref{proc_lwh}) is enhanced when the decay $H^\pm\to h^0W^\pm$ has
a large branching ratio. This occurs at low $\tan\beta$ below the $H^+\to
t\bar b$ kinematic threshold. The cross section falls rapidly as the
$H^+H^-$ pair production becomes unavailable, since there is coupling
suppression as in process (\ref{proc_zwh}).

  Processes (\ref{proc_bwh}) and (\ref{proc_awh}) are not kinematically
enhanced, but they are not coupling suppressed and are large compared to
process (\ref{proc_zwh}). At large $M_{H^\pm}$ the neutral Higgs boson
masses also become large and these two channels are kinematically
suppressed.

  The same situation as process (\ref{proc_zwh}) is found in processes
(\ref{proc_eewh}) -- (\ref{proc_enzh}), shown in figures \ref{rate_eewh}.
The cross sections are small because of the coupling suppression and
because of the cancellation between the $h^0$ and $H^0$ diagrams.

  It is interesting to compare the cross sections of the vector fusion
processes (\ref{proc_enlh}) -- (\ref{proc_enah}), shown in figures
\ref{rate_enlh}, against the $s$-channel processes (\ref{proc_lwh}) --
(\ref{proc_awh}). Although the vector fusion processes are at higher order
in $\alpha_{\rm EW}$, they are not suppressed compared to the $s$-channel
processes especially when the centre-of-mass energy $\sqrt{s}$ is large.
The general behaviour of the cross sections are similar, except the $h^0$
associated production processes (\ref{proc_enlh}) which, compared to
(\ref{proc_lwh}), does not have the enhancement due to the resonant decay
$H^\pm\to h^0W^\pm$. Compared to the $Z^0$ associated production process
(\ref{proc_enzh}), the cross sections are typically two orders of magnitude
higher, when the neutral Higgs masses are small, partly because of the
enhancement coming from the collinear photon which contributes to the
neutral Higgs associated production modes, and partly, again, because of
the cancellation between the $h^0$ and $H^0$ diagrams found in process
(\ref{proc_enzh}).
  The cross sections are small at a 500 GeV collider but the
situation improves at 
a 1 TeV machine, where there are significant regions in $M_{H^\pm}$ and 
$\tan\beta$ where the signal exceeds $10^{-5}$ GeV.

 \section{Discussion of phenomenology}

  We note that processes (\ref{proc_tb}), (\ref{proc_lwh}),
(\ref{proc_bwh}) and (\ref{proc_awh}), as well as the $Z\to b\bar b$ decay
mode of process (\ref{proc_zwh}), which has branching ratio of 15.13\%
\cite{PDG}, all typically lead to the final state $b\bar bH^\pm W^\mp$,
and the different resonance structures imply that there is little
interference between these processes. This indicates that, as proposed
in \cite{ee_bbWH}, $e^-e^+\to b\bar bH^\pm W^\mp$ is an excellent
alternative mode for charged Higgs study at LC, even in cases where the
charged Higgs mass is more than half the collider energy.

  The exact procedure for the tagging of the final state $b\bar bH^\pm
W^\mp$ requires background simulations and is beyond the scope of this
study. Here we only outline the procedure.

  In regions where the dominant decay mode of the charged Higgs bosons is
$\tau^-\bar\nu_\tau$, 
we can look at the hadronic decay of the $W$. If we assume
that the initial state radiation is negligible, we can estimate the
four-momentum of the neutrino. The exact reconstruction is not possible as
$\tau^-$ also carries some missing momentum, and it is dependent on the
$\tau^-$ decay spectrum.

 Although there is large background coming from the decay of the top pair
at $O(0.1~\mathrm{pb})$, the reconstruction of the mass of $\tau^-\bar\nu_\tau$
and other kinematic cuts to suppress events that can come from top pair
production should reduce this background to a negligible level. If the
initial state radiation is not negligible the signal selection becomes
more involved, but we note that the $\tau^-\bar\nu_\tau$ decay mode is relevant
mainly when the charged Higgs mass is below the top quark mass, and in
this region a 500 GeV collider suffices, where the initial state radiation
is relatively small.

  In regions where the dominant decay mode of the charged Higgs bosons is
$t\bar b$ (or $b\bar bW^+$), the final state is $b\bar bb\bar bW^+W^-$. We
can select the leptonic
mode of one of the $W$'s, with the other decaying hadronically,
retaining about $4/9$ of the total rate. We then reconstruct the $W$'s. The
analysis from then on is taxing, but the total SM background rate for the
final state $t\bar tb\bar b$ is $O(\alpha_\mathrm {EW}^2\alpha_S^2)$ and
should be small after we introduce a cut to eliminate soft $b$'s.

  The $\tau\nu_\tau$ associated production mode (\ref{proc_tau}) is interesting
for our purposes only in the large $M_{H^\pm}$ region where the pair
production contribution is suppressed and the dominant decay mode is $t\bar
b$. The dominant background contribution is presumably from top quark pair
production followed by the decay of one of the top quarks into
$b\bar\tau\nu_\tau$. The background reduction procedure would rely on $W$
and $t$ mass reconstruction to eliminate events with two top quarks or with
two $W$'s, and naively this would reduce the background by $\mathcal
O(\alpha_{\mathrm EW}^2)$, so that the signal would be visible. Detailed
simulations are nevertheless desirable.

  In the $H^\pm W^\mp$ associated production mode, if the decay mode is
$\tau^-\bar\nu_\tau$, we can select the hadronic decay of the $W$. We can again
find the missing momentum and reconstruct the charged Higgs mass, and by
filtering events where this mass is small we can eliminate the SM $W^+W^-$
background. If the decay mode is $t\bar b$ we can deal with the top pair
background by totally reconstructing the final state and eliminating events
with two top quarks. This is expected to reduce the top pair background by
about $\alpha_\mathrm{EW}$, hence nearly two orders of magnitude. This
should enable the observation of the peak at the charged Higgs mass in some
regions of the parameter space. We note that the situation in this channel
is better than at the hadron collider \cite{lhc_wh} as the background is
smaller and the final state is more clean. Finally,
we note that at fixed centre-of-mass energy the $H^\pm W^\mp$ cross
section peaks at $M_{H^\pm}\sim m_t+m_b$ \cite{zhu_HW} as seen above, and
at fixed $M_{H^\pm}$ the cross section peaks at $\sqrt{s}\sim 2m_t$
\cite{sk_HW}, which is the energy range that is scanned over for the SM top
threshold studies. At the exact cross section peak, the rate is one order
of magnitude greater, if it is kinematically accessible viz $M_{H^\pm}+
M_{W^\pm}<\sqrt{s}$.

  In all of the above cases, the measurement of the tau
\cite{tau_polarisation} or the top \cite{top_polarisation} polarisation
will provide confirmation of the presence of the charged Higgs.

 \section{Conclusion}

  We discussed the single production of charged Higgs bosons at next
generation linear colliders, and found that the cross sections are large
enough to allow us to use these modes to study the properties of the
charged Higgs bosons. Such analyses will complement the usual pair
production process. Above the kinematical bound for the pair production
process, the single production cross sections are also small, although the
signal remains visible in some channels.

  We found that the $\tau\bar\nu_\tau H^+$ channel, the various channels
contributing to the $b\bar bH^\pm W^\mp$ final state, and the loop induced
$H^\pm W^\mp$ channel are all useful modes for studying charged Higgs
phenomenology. The $H^\pm W^\mp$ channel is enhanced at low $\tan\beta$,
whereas the $\tau\bar\nu_\tau H^+$ channel is enhanced at large $\tan\beta$.
These two are the most promising channels for charged Higgs study beyond
the kinematic limit for pair production.
  The $b\bar bH^\pm W^\mp$ channel, which has contributions from the
$t\bar bH^-$ process and from the $(h^0/H^0/A^0) W^\pm H^\mp$ process, is
large at both large and small values of $\tan\beta$, but the various
channels contributing to this final state are kinematically suppressed and
the cross section falls rapidly at large $M_{H^\pm}$. At small
$M_{H^\pm}$, this mode offers an attractive alternative for studying
charged Higgs boson properties, as there are many different channels
contributing to it, which could be distinguished by means of kinematic
cuts. The analysis of this final state can then be used as a test of the
underlying theory.

  Our analysis has been carried out at the production level only and we
have refrained from commenting on the possible `discovery' or even
`detection' contours. The evaluation of these would require a study of the
decay modes, the partially or fully hadronic final states, and the
backgrounds. The concretisation of these numbers would require machine
dependent detector level simulations which will serve to complement our
analysis.

 \subsection*{Acknowledgements}

  Part of SK's work was done when he worked at Institut f\"{u}r
Theoretische Physik, Universit\"{a}t Karlsruhe. SK would also like to thank
Rutherford Appleton Laboratory (RAL)
for financial support during his visits. SM
and KO thank RAL where part of this work was carried out, and special
thanks go to members of the RAL theory group for many discussions.

 \appendix\refstepcounter{section}
 \section*{Appendix}
  The processes (\ref{proc_tau}) and (\ref{proc_tb}), denoted generically
as:
  \begin{equation}
   e^-_1 e^+_2 \to f_3 \bar{f}_4 H^+_5,
  \end{equation}
  were calculated as follows. The charge conjugate process is equivalent
upto gauge violating effects introduced in the treatment of the widths in
the propagators.

  First, we define the helicity-dependent propagators:
  \begin{equation}
   \Delta_i(\lambda_e,\lambda_i)=\frac
   {Q_eQ_i+
    \frac{s\Delta_Z}{\sin^22\theta_W}\eta_e(\lambda_e)\eta_i(\lambda_i)}
    {s-2p_0\cdot p_i+iM_i\Gamma_i}.
  \end{equation}
  $p_0=p_1+p_2$, $s=p_0^2$ is the centre-of-mass energy squared, and
$\Delta_Z$ is the $Z^0$ propagator. The coupling coefficients $Q$ and
$\eta$ are given in table \ref{app_coup}, where we defined
$x_W=2\sin^2\theta_W$ for convenience.

  \TABLE[hbt]{
  \begin{tabular}{|c|c|c|c|c|}                                \hline
            & $e$      & $H^+$   & $b,\tau$    & $t,\nu$    \\\hline
  $Q$       & $-1$     & $1$     & $Q_3$       & $Q_4$      \\
  $\eta(L)$ & $-1+x_W$ & \raisebox{-\height}[0pt][0pt]{$1-x_W$}
                                 & $-1-Q_3x_W$ & $1-Q_4x_W$ \\
  $\eta(R)$ & $x_W$    &         & $-Q_3x_W$   & $-Q_4x_W$  \\\hline
  \end{tabular}
  \caption{The coupling coefficients in processes
  (\ref{proc_tau}) and (\ref{proc_tb}).}\label{app_coup}
  }

  In terms of these propagators the amplitude is written:
  \begin{eqnarray}\nonumber
   \mathcal{M}_{\rm tot}(\lambda_e) &=&
   \frac{e^3}{sM_W\sqrt{x_W}}\times\sum_\lambda \\\nonumber
   && \bar u_3\Bigl[
   (m_3\tan\beta P_L+m_4\cot\beta P_R)(\not\!k_3+m_4)
   \not\!\!E_{\lambda_e}\Delta_4(\lambda_e,\lambda)P_\lambda-
   \\\nonumber&&-\not\!\!E_{\lambda_e}\Delta_3(\lambda_e,\lambda)
   P_\lambda(\not\!k_4-m_3)(m_3\tan\beta P_L+m_4\cot\beta P_R)+\\
   &&+\Delta_5(\lambda_e)(2p_5-p_0)\cdot E_{\lambda_e}
   (m_3\tan\beta P_L+m_4\cot\beta P_R)P_\lambda\Bigr]v_4.
  \end{eqnarray}
  Here $P_{L/R}=P_{-/+}=(1-/+\gamma_5)/2$ are the left and right chirality
projection operators. The electron current $E^\mu_{\lambda_e}$ is defined
as:
  \begin{equation}
   E^\mu_{\lambda_e} = \bar v_2\gamma^\mu P_{\lambda_e} u_1.
  \end{equation}
  We define $\mu(+,-)=m_{4,3}(\tan\beta,\cot\beta)/M_W$, and the following
mass dimension $-1$ variables:
  \begin{eqnarray}
   {\rm P}(\lambda_e,\lambda)&=& -\mu(\bar\lambda)\sqrt{s}
     [\Delta_4(\lambda_e,\bar\lambda)+\Delta_3(\lambda_e,\lambda)]\\
   \nonumber {\rm M}(\lambda_e,\lambda)&=& m_4\mu(\bar\lambda)
     [\Delta_4(\lambda_e,\lambda)-\Delta_4(\lambda_e,\bar\lambda)]+\\
     &+& m_3\mu(\lambda)
     [\Delta_3(\lambda_e,\lambda)-\Delta_3(\lambda_e,\bar\lambda)]\\
   {\rm Q}^\mu(\lambda_e,\lambda)&=& \mu(\lambda)[
      \Delta_3(\lambda_e,\bar\lambda) p_3^\mu-
      \Delta_4(\lambda_e,    \lambda)(p_4^\mu-p_0^\mu)+
      \Delta_5(\lambda_e)(p_5^\mu-\frac{p_0^\mu}{2})].
  \end{eqnarray}
  Here $k_3=p_0-p_4$ and $k_4=p_0-p_3$. We introduced the notation
$\bar\lambda=-\lambda$.
  Let us suppress the index $\lambda_e$ from here on. The dependence on
the incoming electron chirality will be implicit. The total amplitude is
now written as:
  \begin{equation}\label{app_simp}
   \mathcal{M}_{\rm tot}(\lambda_e)=\frac{e^3}{s\sqrt{x_W}}
   \sum_\lambda\bar u_3\left[\not\!\!EP_\lambda\left[{\rm P}_\lambda
   s^{-1/2}\not\!p_0+{\rm M}_\lambda\right]+2{\rm Q}_\lambda\cdot E
   P_\lambda\right]v_4.
  \end{equation}
  The requirement of gauge invariance implies that the above amplitude is
equivalent to the amplitude due to a Goldstone boson if we replace $E$ by
the momentum carried by the gauge boson $p_0$, and set all widths equal to
zero. This condition can be stated in terms of the above variables as
follows:
  \begin{equation}
   2{\rm Q}_\lambda\cdot p_0 + {\rm P}_{\bar\lambda}\sqrt{s} = 0,
  \end{equation}
  \begin{equation}
   {\rm M}_\lambda=
   \left(\frac{\lambda\eta_e(\lambda_e)s\Delta_Z}{\sin^22\theta_W}\right)
   \left[\frac{m_3\mu(    \lambda)}{s-2p_0\cdot p_3}-
         \frac{m_4\mu(\bar\lambda)}{s-2p_0\cdot p_4}\right].
  \end{equation}

  Squaring up equation (\ref{app_simp}) and summing over final state
helicities, but not dividing by 4 for combinations of initial state
chiralities, we obtain:
  \begin{eqnarray}\label{app_res}
  \nonumber {|\mathcal{M}_{\rm tot}|^2}(\lambda_e)&=&
  \frac{4e^6}{s^2x_W}\sum_\lambda{\rm Re}\Bigl[
  4p_{\bar\lambda}\cdot p_3\left(|{\rm P}_\lambda|^2p_{\bar\lambda}
  \cdot p_4+|{\rm M}_\lambda|^2p_\lambda\cdot p_4\right)-\\
  \nonumber &-& 4\sqrt{s}{\rm P}_\lambda\left(m_4p_{\bar\lambda}
  \cdot p_3{\rm M}_\lambda^*+m_3p_{\bar\lambda}\cdot p_4
  {\rm M}_{\bar\lambda}^*\right)+m_3m_4s\left({\rm P}_\lambda
  {\rm P}_{\bar\lambda}^*+{\rm M}_\lambda{\rm M}_{\bar\lambda}^*
  \right)+\\\nonumber &+& 2\left({\rm Q}_{\bar\lambda}\cdot E\right)^*
  \Bigl[
  p_3\cdot p_4{\rm Q}_{\bar\lambda}\cdot E-m_3m_4{\rm Q}_\lambda\cdot E
  + \\ &+&
        \Bigl(m_3{\rm M}_{\bar\lambda}-\frac{2{\rm P}_\lambda}{\sqrt{s}}
        p_{\bar\lambda}\cdot p_3\Bigr)p_4\cdot E -
        \Bigl(m_4{\rm M}_{    \lambda}-\frac{2{\rm P}_\lambda}{\sqrt{s}}
        p_{\bar\lambda}\cdot p_4\Bigr)p_3\cdot E
  \Bigr]\Bigr].
  \end{eqnarray}
  We defined $p_\lambda$ such that $p_{\lambda_e}=p_1$ and
$p_{\bar\lambda_e}=p_2$. We observe that the above expression is symmetric
under the exchange $(3\leftrightarrow4), ({\rm M}_\lambda\leftrightarrow
{\rm M}_{\bar\lambda}), ({\rm Q}_\lambda\leftrightarrow-{\rm Q}_\lambda)$,
as the amplitude is. This can easily be seen by running a charge
conjugation operator through equation (\ref{app_simp}).
  For numerical evaluation we can take:
  \begin{eqnarray}
  p_1&=&\frac{\sqrt{s}}{2}\left(1,0,0, 1\right) \\
  p_2&=&\frac{\sqrt{s}}{2}\left(1,0,0,-1\right).
  \end{eqnarray}
  so that for any 4-vector $A^\mu$ and up to the phase of $E_{\lambda_e}$
we have:
  \begin{eqnarray}
  A\cdot p_0&=& \sqrt{s}A^0\\
  A\cdot p_\lambda&=&
     \frac{\sqrt{s}}{2}\left(A^0-\lambda\lambda_e A^z\right)\\
  A\cdot E_{\lambda_e}&=& \sqrt{s}\left(A^x+i\lambda_eA^y\right).
  \end{eqnarray}
  The evaluation is relatively straightforward, and easily testable using
the gauge invariance and charge conjugation symmetry tests given above.

 \end{document}